\documentclass[11pt]{article}
\usepackage{amsmath,latexsym,graphicx,subfigure,url,wrapfig}
\title{Picture-Hanging Puzzles%
  \thanks{A preliminary version of this paper appeared in \emph{Proceedings
    of the 6th International Conference on Fun with Algorithms}, Venice, 2012.}%
}
\author{%
  Erik D. Demaine%
    \thanks{MIT Computer Science and Artificial Intelligence Laboratory,
      32 Vassar St., Cambridge, MA 02139, USA,
      \protect\url{{edemaine,mdemaine,rivest}@mit.edu}}
\and
  Martin L. Demaine%
    \footnotemark[2]
\and
  Yair N. Minsky%
    \thanks{Department of Mathematics, Yale University,
      10 Hillhouse Ave., New Haven, CT 06520, USA,
      \protect\url{yair.minsky@yale.edu}}
\and
  Joseph S. B. Mitchell%
    \thanks{Department of Applied Mathematics and Statistics,
      State University of New York, Stony Brook, NY 11794-3600, USA,
     \protect\url{jsbm@ams.sunysb.edu}.
     Partially supported by NSF grant CCF-1018388.}
\and
  Ronald L. Rivest%
    \footnotemark[2]
\and
  Mihai P\v{a}tra\c{s}cu%
    \thanks{AT\&T Labs---Research, 180 Park Ave., Florham Park, NJ 07932,
      \protect\url{mip@alum.mit.edu}}
}
\date{}

\usepackage
  [breaklinks,bookmarks,bookmarksnumbered,bookmarksopen,bookmarksopenlevel=2]
  {hyperref}
{\makeatletter \hypersetup{pdftitle={\@title}}}

\newif\ifabstract
\abstracttrue
\abstractfalse
\newif\iffull
\ifabstract \fullfalse \else \fulltrue \fi

\advance \topmargin by -\headheight
\advance \topmargin by -\headsep
\evensidemargin \oddsidemargin
{\makeatletter
 \gdef\xxxmark{%
   \expandafter\ifx\csname @mpargs\endcsname\relax 
     \expandafter\ifx\csname @captype\endcsname\relax 
       \marginpar{xxx}
     \else
       xxx 
     \fi
   \else
     xxx 
   \fi}
 \gdef\xxx{\@ifnextchar[\xxx@lab\xxx@nolab}
 \long\gdef\xxx@lab[#1]#2{{\bf [\xxxmark #2 ---{\sc #1}]}}
 \long\gdef\xxx@nolab#1{{\bf [\xxxmark #1]}}
 \long\gdef\xxx@lab[#1]#2{}\long\gdef\xxx@nolab#1{}%
}

{\makeatletter \gdef\fps@figure{!htbp}}

\let\realbfseries=\bfseries
\def\bfseries{\realbfseries\boldmath}

\newtheorem{theorem}{Theorem}
\newtheorem{lemma}[theorem]{Lemma}

\def\x{\allowbreak x}
\def\NOT#1{#1^{-1}}
\def\AND{\hbox{\sc and}}
\def\boldAND{\hbox{\bfseries\small AND}}
\def\OR{\hbox{\sc or}}
\def\boldOR{\hbox{\bfseries\small OR}}
\def\mNC1{{\rm mNC$^1$}}

\newtheorem{puzzle}{Puzzle}
\let\realpuzzle=\puzzle
\def\puzzle[#1]{\realpuzzle[#1]\rm}

\def\solution#1{\par \textbf{Puzzle #1:}}

\let\epsilon=\varepsilon

\begin{document}
\maketitle


\begin{abstract}
  We show how to hang a picture by wrapping rope around $n$ nails,
  making a polynomial number of twists, such that the picture falls
  whenever any $k$ out of the $n$ nails get removed, and the picture
  remains hanging when fewer than $k$ nails get removed.
  This construction makes for some fun mathematical magic performances.
  More generally, we characterize the possible Boolean functions
  characterizing when the picture falls in terms of which nails get removed
  as all monotone Boolean functions.  This construction requires an
  exponential number of twists in the worst case, but exponential
  complexity is almost always necessary for general functions.
\end{abstract}

\section{Introduction}

\begin{figure}
  \centering
  \subfigure[A normal hanging. \label{two normal}]
    {\includegraphics[scale=0.4]{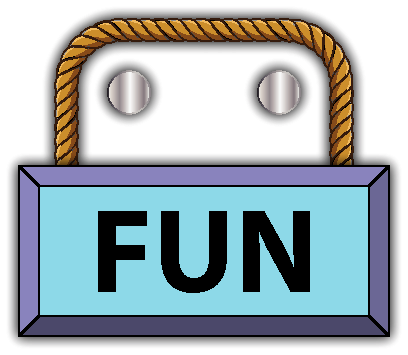}}\hfil\hfil
  \subfigure[Solution to the two-nail puzzle. \label{two solution}]
    {\includegraphics[scale=0.4]{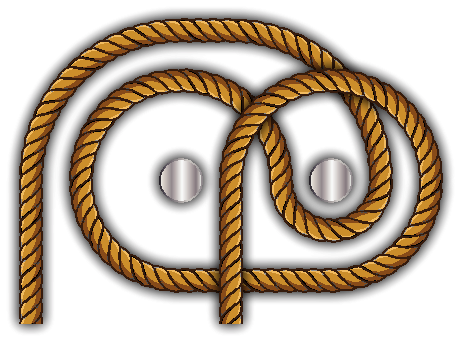}}
  \caption{Two ways to hang a picture by looping around two nails.}
\end{figure}

If you hang a picture with string looped around two nails, and then remove one
of the nails, the picture still hangs around the other nail.  Right?
This conclusion is correct if you hang the picture around the two nails
in the obvious way shown in Figure \ref{two normal}.
An intriguing puzzle, originally posed by A. Spivak in
1997 \cite{Quantum}, asks for a different hanging of the picture
with the property that removing \emph{either} nail causes the picture to fall.
Figure~\ref{two solution} shows a solution to this puzzle.

This puzzle has since circulated around the puzzle community.
Michael Hardy from Harvard posed the puzzle to Marilyn vos Savant
(famous for her claimed ability to answer any riddle),
and the puzzle and solution appeared in her column
\cite{ParadeMagazine}.
Torsten Sillke \cite{Sillke} distributed the puzzle, in particular to
Ed Pegg Jr., and mentioned a connection to Borromean rings
and Brunnian links described in Section~\ref{Borromean Brunnian}.
This connection provides a solution to a more general form of the puzzle,
which we call \emph{$1$-out-of-$n$}: hang a picture on $n$ nails
so that removing any one nail fells the picture.
Pegg's MathPuzzle.com \cite{MathPuzzle} has facilitated a discussion between
Sillke, Neil Fitzgerald, and Chris Lusby Taylor.
Fitzgerald pointed out a connection to group theory,
described in Section~\ref{free group},
which provides a direct solution to the $1$-out-of-$n$ puzzle.
Taylor pointed out a more efficient solution to the same puzzle.
All of this work is detailed and carefully analyzed in Section~\ref{theory}.

We consider a more general form of the puzzle where we want the removal
of certain subsets of nails to fell the picture.  We show that any such
puzzle has a solution: for any collection of subsets of nails,
we can construct a picture hanging that falls when any entire subset of
nails gets removed, but remains hanging when every subset still has at
least one unremoved nail.  This result generalizes picture-hanging puzzles
to the maximum extent possible.

Unfortunately, our construction makes an exponential number of twists
around the $n$ nails.  Indeed, we show that this is necessary, for most
general settings of the problem.  Fortunately, we find polynomial constructions
for the $1$-out-of-$n$ puzzle, as well as the \emph{$k$-out-of-$n$}
generalization where the picture falls only after removing any
$k$ out of the $n$ nails.  More generally, we show that any monotone
Boolean function in the complexity class \mNC1\ (monotone logarithmic-depth
bounded-fanin circuits) has a polynomial-length solution,
which can also be found by a polynomial-time algorithm.

\begin{figure}
  \subfigure[A solution to Puzzle~\ref{1outta3} implemented by wrapping
             rope around children's arms for the Porter Public Lecture during
             the Joint Mathematics Meetings in Boston, Massachusetts in
             January 2012.]
    {\includegraphics[width=0.49\linewidth]{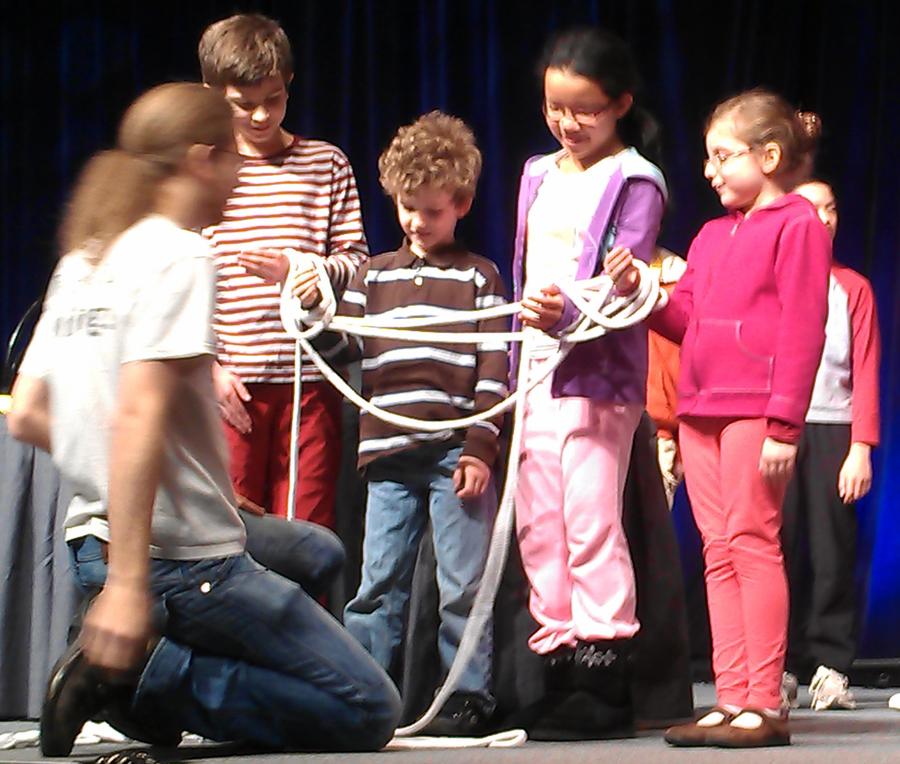}}\hfil\hfil
  \subfigure[A solution to Puzzle~\ref{1or2outta4} implemented by wrapping
             fire hose from the local fire department, when the first author
             forgot to bring his rope for a Poly\'a Lecture in
             Menomonie, Wisconsin in April 2011.]
    {\includegraphics[width=0.49\linewidth]{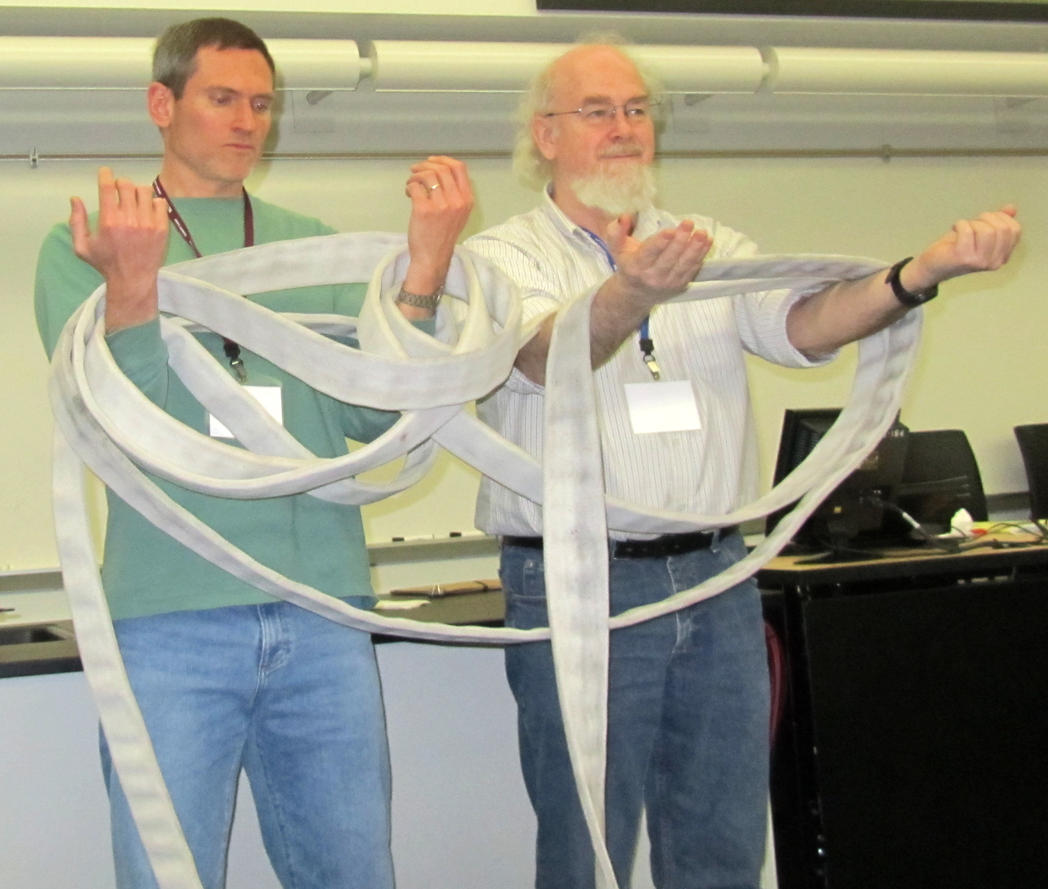}}
  \caption{Picture-hanging puzzles performed as mathematical magic tricks
           by the first author.}
  \label{photos}
\end{figure}

These generalizations make for fun puzzles as well as magic performances.
Section~\ref{puzzles} gives several puzzles accessible to the public
that become increasingly easier to solve while reading through this paper.
These constructions have been featured as a kind of mathematical magic trick
during several of the first authors' talks (first his FUN 2004 plenary talk):
the magician wraps large rope around various volunteers' outstretched arms
(which act as the ``nails''), spectators choose which arms to remove
from the construction, and the magician simply ``applies infinite gravity''
(untangles and pulls on the ends of the rope) to cause the rope to
mathemagically fall to the ground.  Figure~\ref{photos} shows some examples.

Our work interrelates puzzles, magic, topology, Borromean rings,
Brunnian links, group theory, free groups, monotone Boolean function theory,
circuit complexity, AKS sorting networks, combinatorics, and algorithms.

A related result constructs interlocked 2D polygons that separate
(fall apart) when certain subsets of polygons are removed,
again according to an arbitrary monotone Boolean function
\cite{InterlockedPolygons_CCCG2010}.  That result is essentially a
geometric analog of the topological results presented here, although
most of the challenges and remaining open questions differ substantially.

\section{Puzzles}
\label{puzzles}

To whet the appetite of puzzle aficionados, we present a sequence of
picture-hanging puzzles ranging from simple to more interesting extensions,
some of which require rather involved constructions.
We have tested our solutions with $38$-inch lanyard wrapped around fingers,
and found that this length suffices for Puzzles \ref{1outta3}, \ref{2outta3},
\ref{1or2outta3}, \ref{3outta4}, \ref{2or2outta4}, and \ref{1or2outta4},
but for the other puzzles you would need a longer cord or string.
In public performances with large rope wrapped around volunteers' arms,
the first author typically performs Puzzles~\ref{1outta3}, \ref{1outta4},
\ref{2outta3}, \ref{3outta4}, and~\ref{1or2outta4}.

\begin{puzzle}[1-out-of-3] \label{1outta3}
  Hang a picture on three nails so that removing any one nail fells the
  picture.
\end{puzzle}

\begin{puzzle}[2-out-of-3] \label{2outta3}
  Hang a picture on three nails so that removing any two nails fells the
  picture, but removing any one nail leaves the picture hanging.
\end{puzzle}

\begin{puzzle}[1+2-out-of-3] \label{1or2outta3}
  Hang a picture on three nails so that removing the first nail fells the
  picture, as does removing both the second and third nails, but removing
  just the second or just the third nail leaves the picture hanging.
\end{puzzle}

\begin{puzzle}[1-out-of-4] \label{1outta4}
  Hang a picture on four nails so that removing any one nail fells the
  picture.
\end{puzzle}

\begin{puzzle}[2-out-of-4] \label{2outta4}
  Hang a picture on four nails so that removing any two nails fells the
  picture, but removing any one nail leaves the picture hanging.
\end{puzzle}

\begin{puzzle}[3-out-of-4] \label{3outta4}
  Hang a picture on four nails so that removing any three nails fells the
  picture, but removing just one or two nails leaves the picture hanging.
\end{puzzle}

\begin{puzzle}[2+2-out-of-2+2] \label{2or2outta4}
  Hang a picture on two red nails and two blue nails so that
  removing both red nails fells the picture,
  as does removing both blue nails,
  but removing one nail of each color leaves the picture hanging.
\end{puzzle}

\begin{puzzle}[1+2-out-of-2+2] \label{1or2outta4}
  Hang a picture on two red nails and two blue nails so that
  removing any one red nail fells the picture,
  as does removing both blue nails,
  but removing just one blue nail leaves the picture hanging.
\end{puzzle}

\begin{puzzle}[1+3-out-of-3+3] \label{1or3outta6}
  Hang a picture on three red nails and three blue nails so that
  removing any one red nail fells the picture, as does removing all three
  blue nails, but removing just one or two blue nails leaves the picture
  hanging.
\end{puzzle}

\begin{puzzle}[1+2-out-of-3+3] \label{1or2outta6}
  Hang a picture on three red nails and three blue nails so that
  removing any one red nail fells the picture, as does removing any two
  of the blue nails, but removing just one blue nail leaves the picture
  hanging.
\end{puzzle}


\begin{puzzle}[1+1-out-of-2+2+2] \label{two1soutta6}
  Hang a picture on two red nails, two green nails, and two blue nails
  so that removing two nails of different colors (one red and one green,
  or one red and one blue, or one green and one blue) fells the picture,
  but removing two nails of the same color leaves the picture hanging.
\end{puzzle}

\section{Basic Theory: $1$-out-of-$n$}
\label{theory}

We start our mathematical and algorithmic study of picture-hanging puzzles
with the simplest generalization, called \emph{$1$-out-of-$n$}, where
the goal is to hang a picture on $n$ nails such that removing any one nail
fells the picture.  This generalization is what has been studied in the past.
Our contribution is to give a thorough complexity analysis of the resulting
solutions, the best of which Theorem~\ref{basic} summarizes below.

Then, in Section~\ref{disjoint}, we give a slight generalization to handle
colored nails, which is enough to solve many of the puzzles listed above.


\subsection{Connection to Borromean and Brunnian Links}
\label{Borromean Brunnian}

According to Torsten Sillke \cite{Sillke}, Werner Schw\"arzler observed that
the Borromean rings provide a solution to the two-nail picture-hanging problem,
and that generalized forms of Borromean rings provide solutions to more general
picture-hanging problems.  This section describes these and additional
connections.

\begin{wrapfigure}{r}{1.5in}
  \centering
  \includegraphics[scale=0.4]{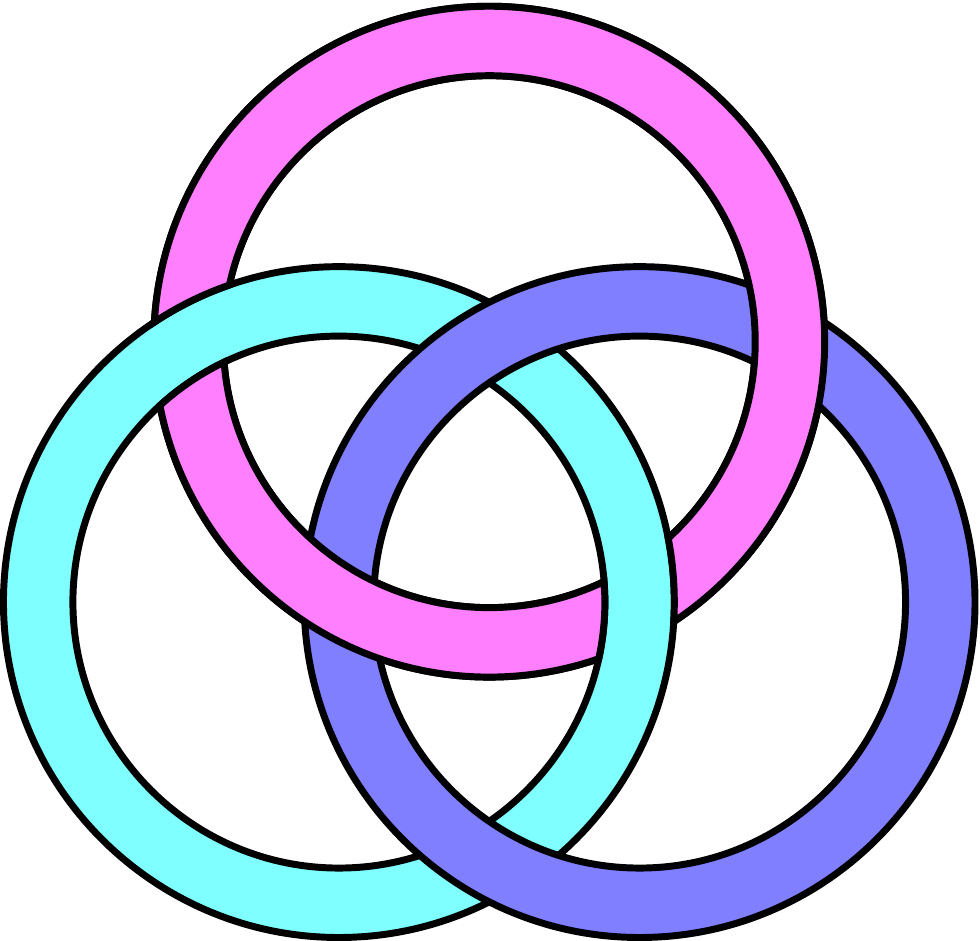}
  \caption{The Borromean rings.}
  \label{Borromean}
\end{wrapfigure}

The classic \emph{Borromean rings} are three loops that are inseparable---in
topology terms, \emph{nontrivially linked}---but such that no two of the rings
are themselves linked.
Figure~\ref{Borromean} shows the standard drawing as interwoven circles,
used by the Italian Renaissance family Borromeo as their family crest.

The property of Borromean rings sounds similar to the picture-hanging puzzle:
the three loops are linked, but removing any one loop unlinks them.
Indeed, by stretching one loop to bring a point to infinity, 
and straightening out the loop, we can view a loop as a
line---or nail---that penetrates the entire construction.
Applying this topology-preserving transformation to two out of the three loops,
we convert any Borromean-ring construction into a solution
to the two-nail picture-hanging puzzle.
Conversely, any solution to the two-nail picture-hanging puzzle
can be converted into a Borromean-ring construction by
viewing the nails as lines piercing the loop of rope
and converting these lines to large loops.

Knot theorists have studied two generalizations to the Borromean rings.
The first generalization, a \emph{Borromean link}, is a collection of $n$ loops
that are linked but such that no two of the loops are linked.
This property seems less useful for an $n$-nail picture-hanging puzzle,
because it guarantees only that removing $n-2$ of the nails fells the picture;
removing between $1$ and $n-3$ of the nails might fell the picture or might
not, depending on the particular Borromean link at hand.
The second generalization, a \emph{Brunnian link}, is a collection of $n$ loops
that are linked but such that the removal of any loop unlinks the rest.
This property is exactly what we need for the $n$-nail picture-hanging puzzle
where removing any one of the $n$ nails fells the picture.
Figure \ref{Brunnian} shows an example of transforming a Brunnian link
into a picture-hanging puzzle.

\iffull
\begin{figure}
  \centering
  \subfigure[Brunnian $4$-link.]
    {\includegraphics[scale=0.4]{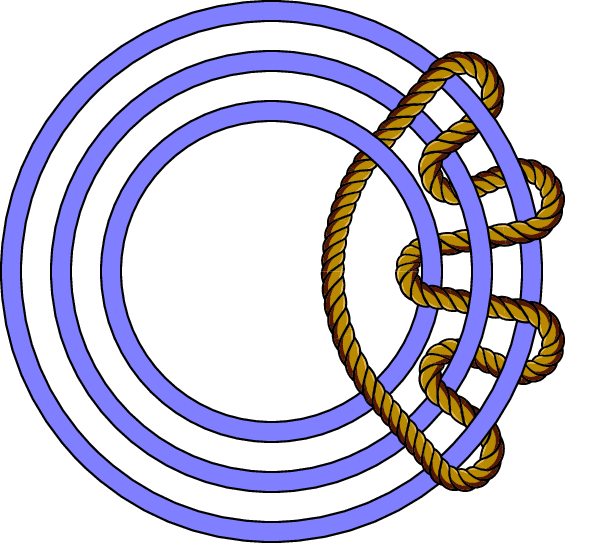}}\hfil\hfil
  \subfigure[Stretching all but one loop into lines.]
    {\includegraphics[scale=0.4]{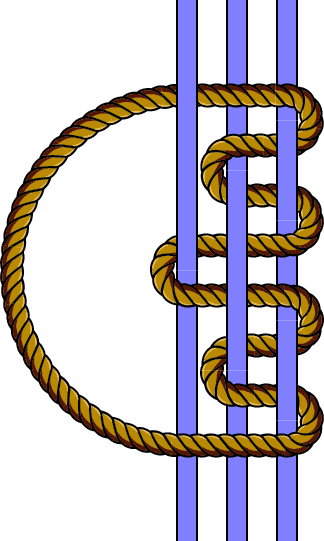}}

  \subfigure[Picture-hanging equivalent from over-under pattern of~(b).]
    {\includegraphics[scale=0.4]{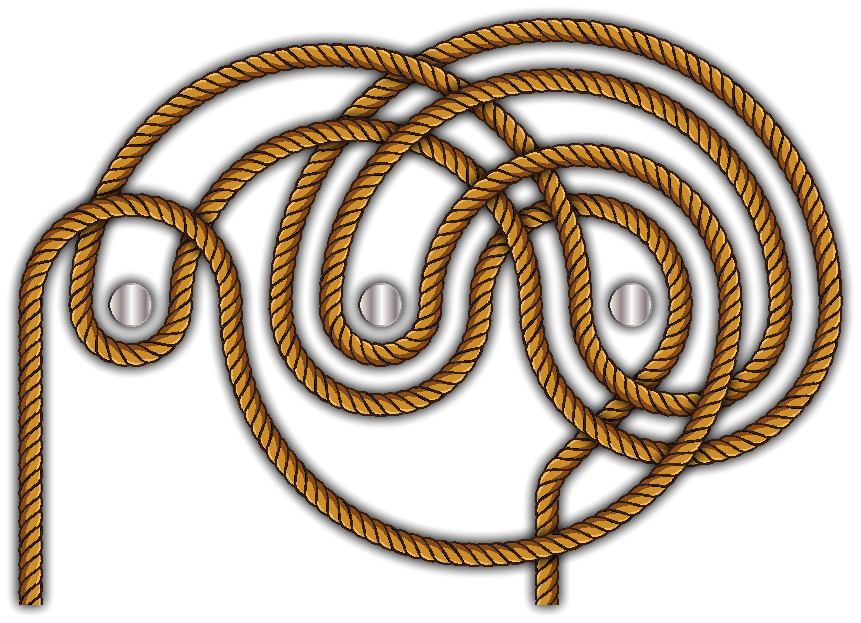}}
  \caption{Transforming a Brunnian $n$-link into a $1$-out-of-$n$
    picture-hanging puzzle, for $n=4$.}
  \label{Brunnian}
\end{figure}
\else
\begin{figure}
  \centering
  \subfigure[Brunnian $4$-link.]
    {\includegraphics[scale=0.25]{brunnian_4a_clean}}\hfil\hfil
  \subfigure[Stretching all but one loop into lines.]
    {\includegraphics[scale=0.25]{brunnian_4b_clean}}\hfil\hfil
  \subfigure[Picture-hanging equivalent from over-under pattern of~(b).]
    {\includegraphics[scale=0.25]{brunnian_4c_clean}}
  \caption{Transforming a Brunnian $n$-link into a $1$-out-of-$n$
    picture-hanging puzzle, for $n=4$.}
  \label{Brunnian}
\end{figure}
\fi

Hermann Brunn \cite{Brunn-1892} introduced Brunnian links in 1892,
about 25 years after the first mathematical study
of Borromean links \cite{Tait-1876}.
Brunn gave a construction for a Brunnian link of $n$ loops
for every $n \geq 3$.
See \cite{Rolfsen-1976} for a more accessible description of this construction.
Using the reduction just described, we obtain a solution to the $1$-out-of-$n$
picture-hanging puzzle for any $n \geq 2$.
The only negative aspect of this solution is that its
``length''---combinatorial complexity, such as the number of crossings in
Figure~\ref{Brunnian}(b)---grows exponentially with~$n$;
we will see a better solution in Section \ref{1outtan}.


Theodore Stanford \cite{Stanford} studied a generalized form of
Brunnian links, or equivalently (similar to Figure~\ref{Brunnian}(a--b)),
Brunnian \emph{braids}.  Specifically, he characterized which braids trivialize
(fall apart) after the removal of all strands in set $S_1$, or all
strands in set $S_2$, etc., for $k$ given sets $S_1, S_2, \dots, S_k$.
This specification might seem equivalent
to an arbitrary monotone Boolean formula,
by writing the formula in disjunctive normal form and setting each $S_i$ to
the terms in the clause.  However, Stanford's characterization of solutions to
this braid problem does not solve the picture-hanging problem because of a key
but subtle difference.  (Indeed, for years, we missed the difference, and
thought that the general picture-hanging problem had already been solved.)
Namely, the braids are permitted to trivialize (fall apart) in other
situations, even when no entire set $S_i$ gets removed.  For example,
the trivial braid is considered a solution, even though no strands need to
be removed.  Less trivially, any $1$-out-of-$n$ braid (where the braid falls
apart from the removal of any strand) would be considered a solution to the
Stanford braid problem.  By contrast, in the picture-hanging puzzle, we want
the picture to fall \emph{exactly when} an entire set $S_i$ gets removed,
remaining intact when only proper subsets have been removed.
As far as we can tell, Stanford's characterization does not lead to a
construction of such a solution, as his proof is concerned mainly with
showing that solutions have a specific form, not with actually constructing
solutions.

\subsection{Connection to Free Group}
\label{free group}

This section describes a more general framework to study picture-hanging
puzzles in general.  The framework is based on group theory and comes naturally
from algebraic topology.  To the best of our knowledge, this connection was
first observed by Neil Fitzgerald \cite{MathPuzzle}.  Although we do not
justify here why the group-theoretic representation is accurate, this is an
easy exercise for those familiar with algebraic topology.

A powerful way to abstract a weaving of the rope around $n$ nails
uses what is called the \emph{free group on $n$ generators}.
Specifically, we define $2 n$ symbols:

\begin{wrapfigure}{r}{2.5in}
  \centering
  \vspace{-3ex}
  \includegraphics[scale=0.4]{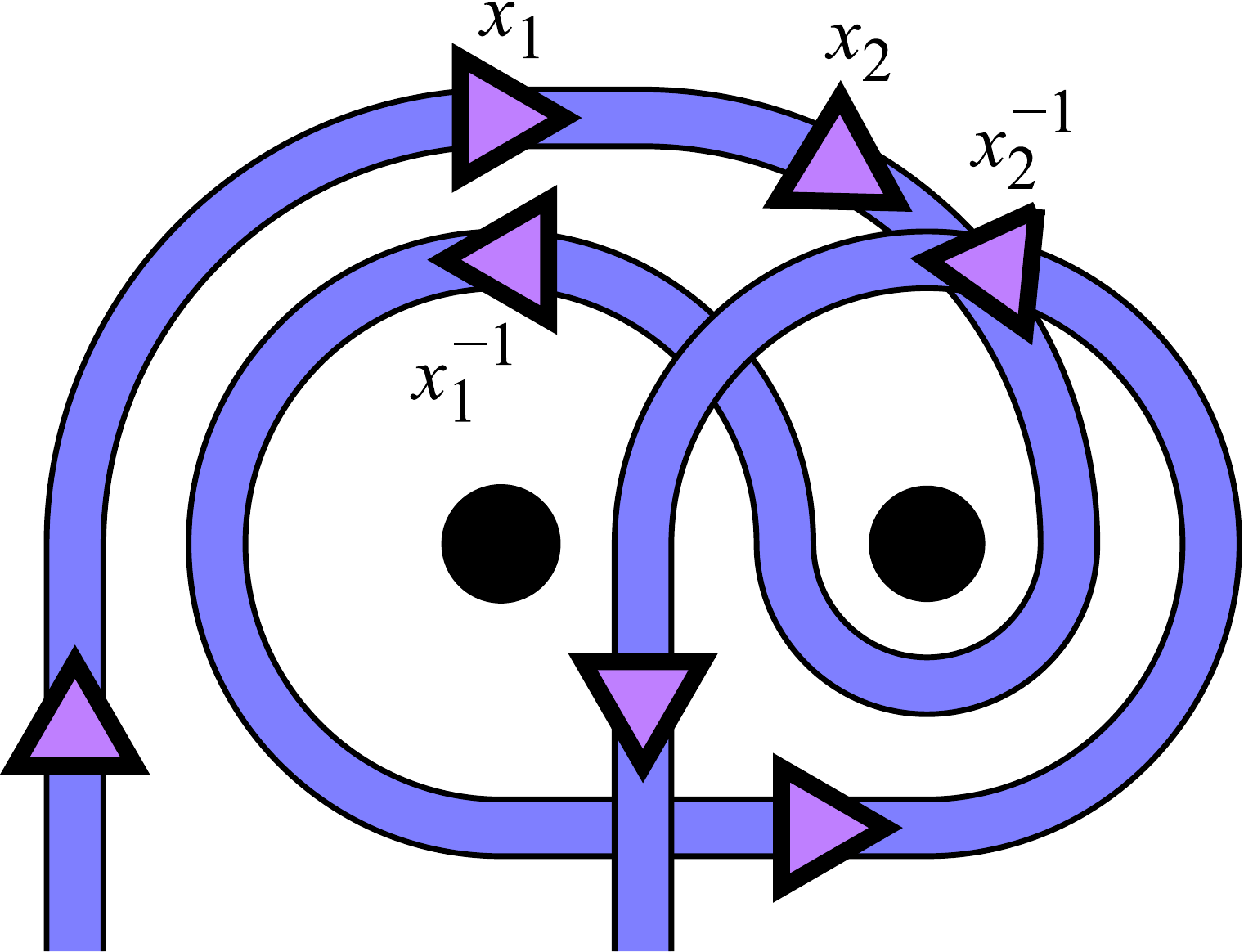}
  \caption{Understanding the algebraic notation for
    Figure~\protect\ref{two solution}.}
  \label{two notation}
\end{wrapfigure}
$$ x_1,\ \NOT{x_1},\ x_2,\ \NOT{x_2},\ \dots,\ x_n,\ \NOT{x_n}. $$
Each $x_i$ symbol represents wrapping the rope around the $i$th nail
clockwise, and each $\NOT{x_i}$ symbol represents wrapping the rope around
the $i$th nail counterclockwise.
Now a weaving of the rope can be represented by a sequence of
these symbols.  For example, the solution to the two-nail picture-hanging
puzzle shown in Figure \ref{two notation}
can be written $x_1 x_2 \NOT{x_1} \NOT{x_2}$ because, starting from the left,
it first turns clockwise around the first (left) nail, then turns clockwise
around the second (right) nail, then turns counterclockwise around the first
nail, and finally turns counterclockwise around the second nail.

In this representation, removing the $i$th nail corresponds to dropping
all occurrences of $x_i$ and $\NOT{x_i}$ in the sequence.
Now we can see why Figure \ref{two notation} disentangles when we
remove either nail.  For example, removing the first nail leaves
just $x_2 \NOT{x_2}$, i.e., turning clockwise around the second nail
and then immediately undoing that turn by
turning counterclockwise around the same nail.
In general $x_i$ and $\NOT{x_i}$ cancel, so all occurrences of $x_i \NOT{x_i}$
and $\NOT{x_i} x_i$ can be dropped.
(The free group specifies that these cancellations are \emph{all}
the simplifications that can be made.)
Thus the original weaving $x_1 x_2 \NOT{x_1} \NOT{x_2}$ is nontrivially
linked with the nails because nothing simplifies;
but if we remove either nail, everything cancels and
we are left with the empty sequence, which represents the trivial weaving
that is not linked with the nails (i.e., the picture falls).

In group theory, the expression $x_1 x_2 \NOT{x_1} \NOT{x_2}$ is called the
\emph{commutator} of $x_1$ and $x_2$, and is written $[x_1, x_2]$.
The commutator is a useful tool for solving more general
picture-hanging puzzles.

\paragraph{Terminology.}
\label{terminology}

In general, define a \emph{picture hanging on $n$ nails} to be a word
(sequence of symbols) in the free group on $n$ generators.
We refer to the number of symbols in the word as the \emph{length} of
the hanging, as it approximates the needed length of the string or cord.
The special identity word $1$ represents the \emph{fallen} state.
\emph{Removing} the $i$th nail corresponds to removing all occurrences
of $x_i$ and $\NOT{x_i}$, which may or may not cause the hanging to fall.

\subsection{$1$-out-of-$n$}
\label{1outtan}

We can use the free-group representation just described to solve the
$1$-out-of-$n$ picture-hanging puzzle.

\begin{theorem} \label{basic}
  For any $n \geq 1$, there is a picture hanging on $n$ nails of length
  at most $2 n^2$ that falls upon the removal of any one nail.
  For each $i = 1, 2, \dots, n$,
  symbols $x_i$ and $\NOT{x_i}$ appear at most $2 n$ times.
\end{theorem}


\ifabstract
\begin{wrapfigure}{r}{3in}
  \centering
  \includegraphics[scale=0.24]{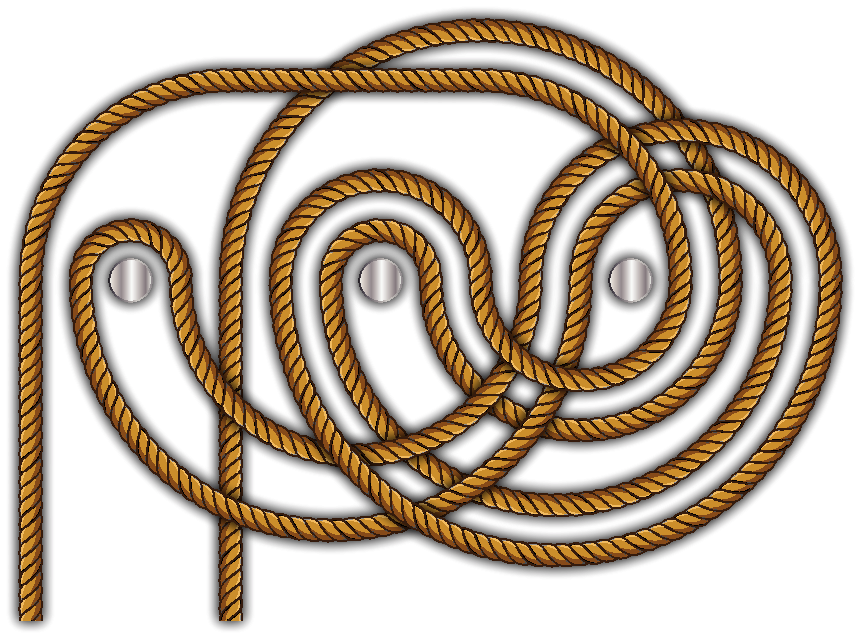}
  \caption{Hanging a picture on three nails so that removing any one nail
           fells the picture.}
  \label{1outta3 fig}
\end{wrapfigure}
\fi

\paragraph{Exponential construction.}

We start with a simpler, less-efficient construction
given by Neil Fitzgerald \cite{MathPuzzle}.%
\footnote{%
  This construction also turns out to be essentially the same as the solution
  that comes out of the Brunnian-link construction described in
  Section \ref{Borromean Brunnian}.
}
The idea is to generalize the weaving
$x_1 x_2 \NOT{x_1} \NOT{x_2}$ by replacing each $x_i$ with an inductive
solution to a smaller version of the problem.
In other words, we start with the solution for $n=2$:
\iffull$$\else$ \fi
  S_2 = [x_1, x_2] = x_1 x_2 \NOT{x_1} \NOT{x_2}.
\iffull$$\else$ \fi
Now from this solution $S_2$ we build a solution for $n=3$ by using the same
pattern but involving copies of $S_2$ in place of one of the $x_i$'s:
\iffull
\begin{eqnarray*}
  S_3 &=& [S_2, x_3] \\
      &=& S_2 x_3 \NOT{S_2} \NOT{x_3} \\
      &=& (x_1 x_2 \NOT{x_1} \NOT{x_2}) x_3
          \NOT{(x_1 x_2 \NOT{x_1} \NOT{x_2})} \NOT{x_3} \\
      &=& x_1 x_2 \NOT{x_1} \NOT{x_2} x_3
          x_2 x_1 \NOT{x_2} \NOT{x_1} \NOT{x_3}.
\end{eqnarray*}
\else
 $S_3 = [S_2, \x_3]
      = S_2 \x_3 \NOT{S_2} \NOT{\x_3}
      = (\x_1 \x_2 \NOT{\x_1} \NOT{\x_2}) \x_3
        \NOT{(\x_1 \x_2 \NOT{\x_1} \NOT{\x_2})} \NOT{\x_3}
      = \x_1 \x_2 \NOT{\x_1} \NOT{\x_2} \x_3
        \x_2 \x_1 \NOT{\x_2} \NOT{\x_1} \NOT{\x_3}$.
\fi
Here we are using the algebraic rules
$\NOT{(x y)} = \NOT{y} \NOT{x}$ and $\NOT{(\NOT{x})} = x$.
Figure~\ref{1outta3 fig} shows the actual picture-hanging solution
corresponding to this sequence.

\iffull
\begin{figure}
  \centering
  \includegraphics[scale=0.4]{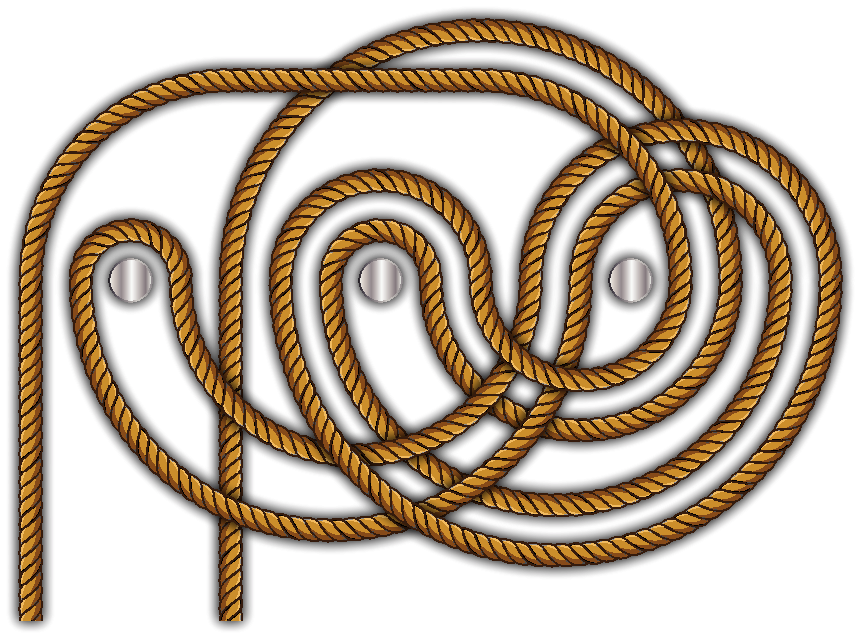}
  \caption{Hanging a picture on three nails so that removing any one nail
           fells the picture.}
  \label{1outta3 fig}
\end{figure}
\fi

\iffull
Granted, this $n=3$ solution $S_3$ is a bit complicated, but we can nonetheless
verify that it satisfies the desired properties.  First, as written, nothing
cancels, so it holds the picture without removing any nails.  Second, if we
remove the first nail $x_1$, then each of the two copies of $S_2$ disappear
because of an $x_2 \NOT{x_2}$ cancellation, so we are left with just
$x_3 \NOT{x_3}$ which also disappears.
Similarly, if we remove the second nail $x_2$, again the copies of $S_2$
collapse and so the entire expression disappears.
Finally, if we remove the third nail $x_3$, then we are left with
$$S_2 \NOT{S_2} = x_1 x_2 \NOT{x_1} \NOT{x_2}
                  x_2 x_1 \NOT{x_2} \NOT{x_1}, $$
in which again everything cancels.
Therefore, no matter which nail we remove, the picture falls.
\fi

Naturally, this construction generalizes to all $n$ by defining
$S_n = [S_{n-1}, x_n] = S_{n-1} x_n \NOT{S_{n-1}} \NOT{x_n}$.
For example,
\iffull
\begin{eqnarray*}
  S_4 &=& [S_3, x_4] \\
      &=& S_3 x_4 \NOT{S_3} \NOT{x_4} \\
      &=& x_1 x_2 \NOT{x_1} \NOT{x_2} x_3
          x_2 x_1 \NOT{x_2} \NOT{x_1} \NOT{x_3} x_4
          x_3 x_1 x_2 \NOT{x_1} \NOT{x_2} \NOT{x_3} x_2 x_1 \NOT{x_2} \NOT{x_1}
          \NOT{x_4}.
\end{eqnarray*}
\else
 $S_4 = [S_3, \x_4]
      = S_3 \x_4 \NOT{S_3} \NOT{\x_4}
      = \x_1 \x_2 \NOT{x_1} \NOT{\x_2} \x_3
        \x_2 \x_1 \NOT{x_2} \NOT{\x_1} \NOT{\x_3} \x_4
        \x_3 \x_1 \x_2 \NOT{\x_1} \NOT{\x_2} \NOT{\x_3} \x_2 \x_1 \NOT{\x_2} \NOT{\x_1}
        \NOT{\x_4}$.
\fi
By induction, if we remove any of the first three nails, the two copies of
$S_3$ disappear, leaving us with $x_4 \NOT{x_4}$ which cancels.
And if we remove the fourth nail $x_4$, we are left with $S_3 \NOT{S_3}$
which cancels.

The problem with this construction, which we start to see with the full
expansion of $S_4$, is that the length of the sequence $S_n$
grows exponentially with~$n$.  More precisely, the number of symbols in $S_n$
is $2^n + 2^{n-1} - 2$.
To see why this count is correct,
first check that $S_2$ has length $4 = 2^2 + 2^1 - 2$.
Then,
if we suppose inductively that $S_{n-1}$ has length $2^{n-1} + 2^{n-2} - 2$,
we can conclude that $S_n$ has twice that length plus $2$ for the occurrences
of $x_n$ and $\NOT{x_n}$, for a total of
\iffull
\begin{eqnarray*}
  & & 2 (2^{n-1} + 2^{n-2} - 2) + 2 \\
  &=& 2^n + 2^{n-1} - 4 + 2 \\
  &=& 2^n + 2^{n-1} - 2,
\end{eqnarray*}
\else
$2 (2^{n-1} + 2^{n-2} - 2) + 2 = 2^n + 2^{n-1} - 4 + 2 = 2^n + 2^{n-1} - 2$,
\fi
as claimed.

\paragraph{Polynomial construction.}
\label{Polynomial construction}

Fortunately, there is a more efficient construction that solves the
$1$-out-of-$n$ picture-hanging puzzle, which will prove Theorem~\ref{basic}.
This construction was designed by Chris Lusby Taylor \cite{MathPuzzle}.
The idea is to recursively build
$S_n$ in a more balanced way, in terms of $S_{n/2}$ for the first half of the
nails and $S_{n/2}$ for the second half of the nails,
instead of one $S_{n-1}$ and a single variable.
To enable this construction, we need to consider a more general problem
involving the nails from $i$ through $j$ for various $i$ and $j$.
At the simplest level we have a single nail:
\iffull$$\else$ \fi
  E(i:i) = x_i.
\iffull$$\else$ \fi
At the next simplest level we have two nails as before:
\iffull$$\else$ \fi
  E(i:i+1) = [x_i, x_{i+1}] = x_i x_{i+1} \NOT{x_i} \NOT{x_{i+1}}.
\iffull$$\else$ \fi
Then for an arbitrary interval $i : j$, we build $E(i:j)$
out of a recursive copy of $E$ applied to the first half of the interval
and a recursive copy of $E$ applied to the second half of the interval:
\begin{eqnarray*}
  E(i:j) &=& \textstyle
             \left[E\left(i : \left\lfloor {i+j \over 2} \right\rfloor\right),
              E\left(\left\lfloor {i+j \over 2} \right\rfloor+1 : j\right)\right].
\end{eqnarray*}

For $n=3$, this construction does not save anything, because splitting
an interval of length three in half leaves one piece of length two and
one piece of length one.  But for $n=4$ we gain some efficiency:
\begin{eqnarray*}
  E(1:4) &=& [E(1:2), E(3:4)] \\
         &=& E(1:2) \  E(3:4) \  \NOT{E(1:2)} \  \NOT{E(3:4)} \\
         &=& (x_1 x_2 \NOT{x_1} \NOT{x_2})
             (x_3 x_4 \NOT{x_3} \NOT{x_4})
             \NOT{(x_1 x_2 \NOT{x_1} \NOT{x_2})}
             \NOT{(x_3 x_4 \NOT{x_3} \NOT{x_4})} \\
         &=& x_1 x_2 \NOT{x_1} \NOT{x_2}
             x_3 x_4 \NOT{x_3} \NOT{x_4}
             x_2 x_1 \NOT{x_2} \NOT{x_1}
             x_4 x_3 \NOT{x_4} \NOT{x_3}.
\end{eqnarray*}
This sequence has $16$ symbols compared to the $22$ from $S(4)$ above.

This savings becomes substantially more impressive as $n$ grows.
If $n$ is a power of two,
then $E(1:n)$ has length $n^2$, because it consists of two copies of
$E(1:n/2)$ and two copies of $E(n/2+1:n)$ and because $4 (n/2)^2 = n^2$.
Furthermore, in this case,
symbols $x_i$ and $\NOT{x_i}$ appear exactly $n$ times in $E(1:n)$
because by induction they appear exactly $n/2$ times in exactly one of
$E(1:n/2)$ and $E(n/2+1:n)$.

If $n$ is not a power of two, we still have that
$E(1:n)$ has length at most $(2 n)^2 = 4 n^2$,
because $E(1:n)$ only increases if we round up to the next power of two.
The integer sequence formed by the length of $E(1:n)$ with $n=1,2,3,\dots$
is in fact in Neil Sloane's Encyclopedia \cite{Sloane}.
Ellul, Krawetz, Shallit, and Wang \cite{Shallit} proved that,
if $n$ is $b$ larger than the previous power of two, $2^a$,
then the length of $E(1:n)$ is precisely $(2^a)^2 + b (2^{a+2} - 2^a)$.
This formula is always at most $2 n^2$.
Furthermore, symbols $x_i$ and $\NOT{x_i}$ appear at most $2n$ times in
$E(1:n)$ because each recursion doubles the number of appearances,
and there are precisely $\lceil \log_2 n \rceil \leq \log_2 n + 1$ recursions,
so the number of appearances is at most $2^{\log_2 n + 1} = 2 n$.

This completes the proof of Theorem~\ref{basic}.

\subsection{Disjoint Subsets of Nails}
\label{disjoint}

One way to state the most general form of a picture-hanging puzzle is the
following:
given arbitrary subsets $S_1$, $S_2$, \dots, $S_k$ of $\{1, 2, \dots, n\}$,
hang a picture on $n$ nails such that removing all the nails in $S_i$
fells the picture, for any $i$ between $1$ and $k$, but removing a set of
nails that does not include an entire $S_i$ leaves the picture hanging.
For example, the $1$-out-of-$n$ puzzle is the special case of
$S_i = \{i\}$ for $i = 1, 2, \dots, n$.
All of the puzzles posed in Section~\ref{puzzles} can be represented
as particular instances of this general puzzle.

As a warmup, we first illustrate how the
theory we have developed so far easily solves the special case in which the
subsets $S_1, S_2, \dots, S_k$ are pairwise disjoint.
This corresponds to the pegs being divided into different color classes,
and the picture falling precisely when an entire color class has been removed.
Many of the puzzles posed in Section~\ref{puzzles} fall into this class.

\begin{theorem} \label{disjoint theorem}
  For any partition of $\{1, 2, \dots, n\}$ into disjoint subsets
  $S_1, S_2, \dots, S_k$, there is a picture hanging on $n$ nails
  of length at most $2 k n$ that falls when removing all nails in $S_i$,
  for any $i$ between $1$ and $k$, but does not fall when keeping
  at least one nail from each~$S_i$.
\end{theorem}

To prove this theorem,
the idea is to replace each subset $S_i$ of nails with a ``supernail''
and then apply the $1$-out-of-$n$ solution to the supernails.
Whenever the solution says to wrap clockwise around a supernail,
we wrap clockwise around each of the nails in the supernail
in a particular order;
when we wrap counterclockwise around the supernail,
we wrap counterclockwise around each of the nails in the reverse order.

More precisely, we represent each subset
$S_i = \{x_{i,1}, x_{i,2}, \dots, x_{i,r_i}\}$
by the sequence $w_i = x_{i,1} x_{i,2} \cdots x_{i,r_i}$.
This sequence $w_i$ collapses to nothing precisely when
that subset has been satisfied, i.e.,
all of the $x_{i,j}$ nails constituting $S_i$ have been removed.
Our goal is therefore to combine $w_1, w_2, \dots, w_k$ so that the entire
sequence collapses precisely when any of the $w_i$'s collapses.

Next we combine $w_1, w_2, \dots, w_k$ using the $E(1:k)$ construction,
where each $x_i$ in $E(1:k)$ is replaced by a~$w_i$.
In other words, we define $W(i:i) = w_i$ and
\begin{eqnarray*}
  W(i:j) &=& \textstyle
             \left[W\left(i : \lfloor \frac{1}{2} (i+j) \rfloor\right),
              W\left(\lfloor \frac{1}{2} (i+j) \rfloor+1 : j\right)\right].
\end{eqnarray*}
The resulting sequence $W(1:n)$ collapses whenever any of the $w_i$'s
collapse, because by induction the left or right half containing $w_i$
collapses, leaving the other half next to its inverse, which collapses.

The last requirement of the solution is that the sequence does not collapse if
every $w_i$ remains intact.  This property follows because no two
of the $w_i$'s share a letter.  Thus any two subconstructions
$W(i:j)$ and $W(j+1:k)$ do not share a letter.  Therefore, none of the sequence
concatenations we make in the construction of $W(1:n)$ could have accidental
cancellation if every $w_i$ keeps at least one letter.

The length of $W(1:n)$ is at most $2 k$ times the total length of the $w_i$'s,
because Theorem~\ref{basic} guarantees that each $w_i$ appears at most
$2 k$ times (counting the negated form $\NOT{w_i}$).
The total length of the $w_i$'s is exactly $n$ because the subsets $S_i$'s form
a partition of $\{1, 2, \dots, n\}$.
Therefore the total length is at most $2 k n$, which in particular is at most
$2 n^2$, completing the proof of Theorem~\ref{disjoint theorem}.

\section{General Theory}
\label{general}

This section develops a general theory for solving the most general form
of the picture-hanging puzzle.  Section~\ref{disjoint} described
one statement of this general form, using subsets, but this turns out to be an
inefficient way to represent even relatively simple problems.  For example,
the $k$-out-of-$n$ puzzle has ${n \choose k}$ subsets of nails that fell
the picture, which is exponential for $k$ between $\epsilon n$ and
$(1-\epsilon) n$.  We therefore turn to a more general representation,
called ``monotone Boolean functions''.  Although our general solution
remains exponential in the worst case, we show in Section~\ref{k-out-of-n}
how this representation allows us to achieve a polynomial solution for
$k$-out-of-$n$ in particular.

\subsection{Connection to Monotone Boolean Functions}

For a given picture hanging $p$ on $n$ nails, define the
\emph{fall function} $f_p(r_1, r_2, \dots, r_n)$, where each $r_i$ is
a Boolean value (true/1 or false/0), to be a Boolean value specifying
whether the hanging $p$ falls after removing all $x_i$'s
corresponding to true $r_i$'s.  
For example, a solution $p$ to the $1$-out-of-$n$ puzzle has the fall function
``is any $r_i$ set to true?'', because setting any $r_i$ to true (i.e.,
removing any $x_i$) causes the construction $p$ to fall.
Using standard notation from logic,
\begin{equation} \label{1-out-of-n spec}
  f_p(r_1,r_2,\dots,r_n) = r_1 \vee r_2 \vee \cdots \vee r_n,
\end{equation}
where $\vee$ represents \OR\ (logical disjunction).

The most general form of picture-hanging puzzle on $n$ nails is the following:
given a desired fall function $f(r_1, r_2, \dots, r_n)$,
find a picture hanging $p$ with that fall function, i.e., with $f_p = f$.
For example, the function in Equation~(\ref{1-out-of-n spec}) is a
specification of the $1$-out-of-$n$ problem.

Not all such puzzles can be solved, however.
Every fall function must satisfy a simple property called \emph{monotonicity}:
if $r_1 \leq r'_1$, $r_2 \leq r'_2$, \dots, and $r_n \leq r'_n$,
then $f(r_1,r_2,\dots,r_n) \leq f(r'_1,r'_2,\dots,r'_n)$.
Here we view the truth values as 0 (false) and 1 (true),
so that false $<$ true.  This condition just says that,
if the hanging falls when removing certain nails given by the $r_i$'s,
and we remove even additional nails as given by the $r'_i$'s,
then the hanging still falls.  A picture hanging cannot ``unfall'' from
removing nails, so monotonicity is a necessary condition on fall functions.
For example, it is impossible to find a picture hanging that falls from
removing any one nail but not from removing more nails.

Monotone Boolean functions are well-studied in combinatorics
(through Dedekind's Problem), computational complexity,
and computational learning theory, among other fields.
It is well-known that they are exactly the functions formed by
combining the variables $r_1, r_2, \dots, r_n$ with the operators
$\AND$ ($\wedge$) and $\OR$ ($\vee$).  (In particular, \textsc{not}
is forbidden.)  We can leverage this existing theory about
monotone Boolean functions to obtain powerful results about picture hanging.

\subsection{Arbitrary Monotone Boolean Functions}

In particular, we establish that monotone Boolean functions are exactly
the fall functions of picture hangings.  We have already argued that
every fall function is monotone; the interesting part here is that
every monotone Boolean function can be represented as the fall function
of a picture hanging.  Our construction is exponential in the worst case,
but as we will see, is efficient in many interesting cases.

\begin{theorem} \label{arbitrary theorem}
  Every monotone Boolean function $f$ on $n$ variables is the fall function
  $f_p$ of a picture hanging $p$ on $n$ nails.  If the function $f$ can be
  computed by a depth-$d$ circuit of two-input $\AND$ and $\OR$ gates,
  then we can construct $p$ to have length $c^d$ for a constant~$c$.
  We can compute such $p$ in time linear in the length of~$p$.
  In particular, for functions $f$ representable by a depth-$O(\log n)$
  circuit of two-input $\AND$ and $\OR$ gates (the complexity class \mNC1),
  there is a polynomial-length picture hanging.
\end{theorem}

Our approach to proving this theorem is to simulate $\AND$ and $\OR$ gates
in a way that allows us to combine them into larger and larger circuits.
The most intuitive version of the construction is when the function $f$
is represented as a monotone Boolean \emph{formula} (as opposed to circuit),
which can be parsed into a tree with the $r_i$'s at the leaves and the
value of $f$ at the root.  As base cases, we can represent the formula
$r_i$ by the picture hanging $x_i$ (or $\NOT{x_i}$), which falls
precisely when the $i$th nail gets removed.  We show next that,
given picture hangings $p$ and $q$ representing two monotone Boolean
functions $f$ and~$g$, we can construct picture hangings $\AND(p,q)$ and
$\OR(p,q)$ representing $f \wedge g$ and $f \vee g$, respectively.
While most intuitively applied up a tree representing a formula,
the same construction applies to a directed acyclic graph
representing a general circuit.

\paragraph{\boldAND.}
Our $\AND$ and $\OR$ constructions build on two known lemmas from
monotone function theory.  We start with $\AND$:

\begin{lemma} \label{and lemma}
  {\rm [A. I. Mal'tsev] \cite[Lemma~3]{Makanin-1985}}
  For any two words $p,q$ in the free group on $x_1, x_2, \dots, x_n$,
  the equation
  \iffull$$\else$ \fi
    p^2 x_1 p^2 x_1^{-1} = (q x_2 q x_2^{-1})^2
  \iffull$$\else$ \fi
  is equivalent to the conjunction $(p=1) \wedge (q=1)$.
\end{lemma}

Because the free group is a group, we can rewrite this equation as
\iffull$$\else$ \fi
  p^2 x_1 p^2 \NOT{x_1} (q x_2 q \NOT{x_2})^{-2} = 1.
\iffull$$\else$ \fi

Lemma~\ref{and lemma} states that this equation holds
if and only if $p=1$ and $q=1$.  Recall that $1$ is the fallen state of
picture hangings.  Thus the left-hand side
\begin{equation} \label{and eq}
  \AND(p,q)
  = p^2 x_1 p^2 \NOT{x_1} (q x_2 q \NOT{x_2})^{-2}
  = p^2 x_1 p^2 \NOT{x_1} x_2 \NOT{q} \NOT{x_2} \NOT{q} x_2 \NOT{q} \NOT{x_2} \NOT{q}
\end{equation}
falls if and only if both $p$ and $q$ fall.
This construction is our desired~$\AND$.

\paragraph{\boldOR.}
We now turn to the $\OR$ construction:

\begin{lemma} \label{or lemma}
  {\rm [G. A. Gurevich] \cite[Lemma~4]{Makanin-1985}}
  For any two words $p,q$ in the free group on $x_1, x_2, \dots, x_n$,
  the conjunction of the four equations
  \iffull$$\else$ \fi
    (p x_1^s p x_1^{-s}) (q x_2^t q x_2^{-t}) =
    (q x_2^t q x_2^{-t}) (p x_1^s p x_1^{-s})
  \iffull
    \quad \textrm{for all } s,t = \pm 1$$
  \else
    $, for all $s,t = \pm 1$,
  \fi
  is equivalent to the disjunction $p=1 \vee q=1$.
\end{lemma}


Right-multiplying by the inverse of the right-hand side (again),
the equations are equivalent to
\iffull$$\else$ \fi
  (p x_1^s p x_1^{-s}) (q x_2^t q x_2^{-t})
  \NOT{(p x_1^s p x_1^{-s})}
  \NOT{(q x_2^t q x_2^{-t})} = 1
  \quad \textrm{for all } s,t = \pm 1.
\iffull$$\else$ \fi
Using commutator notation, the equations become
\iffull$$\else$ \fi
  \big[p x_1^s p x_1^{-s}, q x_2^t q x_2^{-t}\big] = 1
  \quad \textrm{for all } s,t = \pm 1.
\iffull$$\else$ \fi

Lemma~\ref{or lemma} states that these equations all hold
if and only if $p=1$ or $q=1$.  To obtain the conjunction of the four
equations, we apply the $\AND$ construction above:
\begin{align}
  \OR(p,q) = \AND\bigg(
  & \AND\Big( \big[p x_1 p x_1^{-1}, q x_2 q x_2^{-1}\big],
              \big[p x_1 p x_1^{-1}, q x_2^{-1} q x_2\big] \Big),
  \nonumber\\
  &
    \AND\Big( \big[p x_1^{-1} p x_1, q x_2 q x_2^{-1}\big],
              \big[p x_1^{-1} p x_1, q x_2^{-1} q x_2\big] \Big) \bigg).
  \label{or eq}
\end{align}
%
Thus $\OR(p,q)$ falls if and only if either $p$ or $q$ falls.
This construction is our desired~$\OR$.  The $\OR$ formula
expands to 256 $p$ and $q$ terms, and 822 $x_1$ and $x_2$ terms,
for a total of 1,078 terms.

\paragraph{Analysis.}

Now we argue that a circuit of depth $d$ results in a picture hanging
of length at most $c^d$ for a constant~$c$.  The output of the circuit
is the output of some gate, either $\AND$ or $\OR$, which has two inputs.
Each input can be viewed as the output of a subcircuit of the overall
circuit, with smaller depth $d-1$.  The two subcircuits may overlap
(or even be identical), but we treat them as separate by duplicating
any shared gates.  By induction on depth, these subcircuits can be
converted into picture hangings $p$ and $q$ of length at most $c^{d-1}$.
We combine these picture hangings via $\AND(p,q)$ or $\OR(p,q)$,
according to the output gate type, to obtain our desired picture hanging.
The resulting length is at most the maximum length of $p$ and $q$,
which is at most $c^{d-1}$, times the number of terms in Equations
(\ref{and eq}) and (\ref{or eq}) defining $\AND$ and~$\OR$.
Thus setting $c=1{,}078$ suffices.

In the base case, the depth-$0$ circuit has no gates and simply takes the
value of a variable $r_i$, and we use the picture hanging $x_i$,
which has length $1 = c^0$ as needed.

This argument gives a $1{,}078^d$ upper bound on the length of the constructed
picture hanging.  In fact, only 256 of the 1,078 terms in (\ref{or eq}) are
recursive ($p$~or~$q$), so the upper bound is $256^d$ plus lower-order terms.

\paragraph{Running time.}

To compute the picture hanging resulting from this construction
in linear time, we simply need a data structure for representing a
picture hanging that supports concatenation and inversion in constant time.
One such data structure is a doubly linked list, with a bit in each node
to indicate when the orientation and inverted state flips.
Once we construct the data structure representing the final picture hanging,
a single pass through the list can flatten into a typical
representation of a picture hanging as a word in the free group.

This argument completes the proof of Theorem~\ref{arbitrary theorem}.

\subsection{Worst-Case Optimality}

Unfortunately, most monotone Boolean functions require exponential-length
picture hangings:

\begin{theorem} \label{optimality theorem}
  Almost all monotone Boolean functions require
  length-$\Omega(2^n/(\sqrt n \log n))$ picture hangings.
\end{theorem}

This theorem follows from a counting argument, specifically, contrasting the
large number of monotone Boolean functions with the relatively small number of
picture hangings of a given length.

First we demonstrate a large number of monotone Boolean functions,
using a standard argument.
The vectors $(r_1, r_2, \dots, r_n)$ with exactly
$n/2$ $1$'s (and $n/2$ $0$'s) can all have their function values set
independently.  There are ${n \choose n/2}$ such vectors.
Thus there are at least $2^{n \choose n/2}$ monotone Boolean functions
on $n$ variables.

Next we observe that the number of picture hangings of length $\ell$
is at most $(2 n)^\ell$, because there are at most $2 n$ choices for each
symbol in the word.  (The correct number of choices is $2n-1$,
except for the first, to avoid cancelation.)
The number of picture hangings of length at most $\ell$ is
$\sum_{i=1}^\ell (2 n)^i < 2 (2 n)^\ell$.

To represent all monotone Boolean functions, we must have
\iffull$$\else$ \fi
  2 (2 n)^\ell \geq 2^{n \choose n/2}.
\iffull$$\else$ \fi
Taking $\log_2$ of both sides, we must have
\iffull$$\else$ \fi
  1 + \ell (1 + \log_2 n) \geq {n \choose n/2}.
\iffull$$\else$ \fi
Asymptotically, ${n \choose n/2} \sim 2^n \sqrt{2 \over \pi n}$.
Thus we must have $\ell \sim 2^n \sqrt{2 \over \pi n \log_2 n}$
to have all monotone Boolean functions represented by picture hangings
of length~$\ell$.

The number of picture hangings shorter than this length $\ell$ is
asymptotically smaller than the number of picture hangings of length $\ell$
(by a factor of about $1/n$), and thus too small to represent all but a
asymptotically vanishing fraction of monotone Boolean functions.
Therefore, almost every (asymptotically most) picture hangings have length
at least~$\ell$, completing the proof of Theorem~\ref{optimality theorem}.

\subsection{$k$-out-of-$n$}
\label{k-out-of-n}

One example of a picture-hanging puzzle that can be solved more efficiently
(at least in theory) is the $k$-out-of-$n$ puzzle:

\begin{theorem}
  For any $n \geq k \geq 1$, there is a picture hanging on $n$ nails,
  of length $n^{c'}$ for a constant~$c'$,
  that falls upon the removal of any $k$ of the nails.
\end{theorem}

We simply argue that the monotone Boolean function ``are at least $k$ of
the $r_i$'s true?''\ is in the complexity class \mNC1, that is,
can be represented by a logarithmic-depth binary circuit.
The idea is to \emph{sort} the $r_i$ values,
again viewing Boolean values as $0$ (false) and $1$ (true).
The result of this sorting is a sequence of $j$ $0$'s
followed by a sequence of $n-j$ $1$'s.
Our goal is to determine whether $n-j \geq k$.
To do so, we would simply look at the $(n-k+1)$st item in the sorted order:
if it is $1$, then there at least $k$ $1$'s, and otherwise, there are fewer.

Our construction thus starts from a logarithmic-depth \emph{sorting network},
\iffull\xxx{perhaps cite Knuth volume 3, and/or CLRS, for sorting networks in general}\fi
as first achieved by Ajtai, Koml\'os, and Szemer\'edi
\cite{Ajtai-Komlos-Szemeredi-1983} and improved by Mike Paterson
\cite{Paterson-1990}.  A sorting network consists of a circuit of
\emph{comparators}.  Each comparator has two inputs and two outputs,
outputting the smaller input on top and the larger input on bottom
(thus sorting the inputs).

\begin{figure}
  \centering
  \includegraphics[scale=1.2]{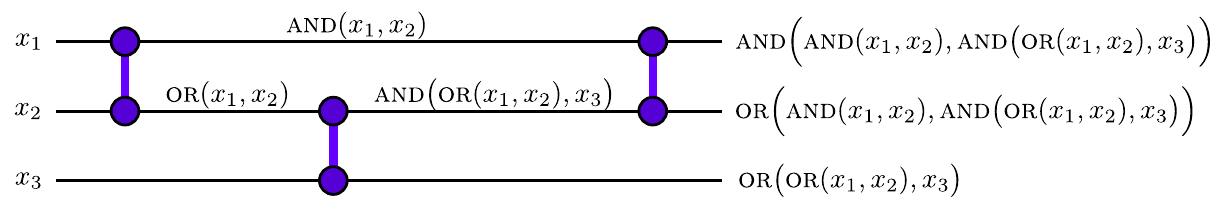}
  \caption{Converting a sorting network (with three elements and three
           comparators) into a sequence of (three) monotone Boolean formulas.}
  \label{sorting network}
\end{figure}

Such a sorting network can be converted into monotone Boolean formulas,
as illustrated in Figure~\ref{sorting network}.  If a comparator has two
inputs $p$ and $q$, then the top (minimum) output is $p \wedge q$,
and the bottom (maximum) output is $p \vee q$.  Applying this rule
to every comparator, we end up with a monotone Boolean formula
(or, more efficiently, a monotone Boolean circuit) representing each
of the items in the sorted output.  The solution to the $k$-out-of-$n$
sorting puzzle is simply the $(n-k+1)$st of these circuits.

Because the sorting network has logarithmic depth, so does the monotone
Boolean circuit; the depths are actually identical.
The best known upper bound on the depth of sorting networks
is under $6{,}100 \log_2 n$.  Applying Theorem~\ref{arbitrary theorem},
we obtain a picture hanging of length
$c^{6{,}100 \log_2 n} = n^{6{,}100 \log_2 c}$.
Using the $c \leq 1{,}078$ upper bound,
we obtain an upper bound of $c' \leq 6{,}575{,}800$.
Using the $c \leq 256 + o(1)$ upper bound,
we obtain an upper bound of $c' \leq 1{,}561{,}600 + o(1)$.

So, while this construction is polynomial, it is a rather large polynomial.
For small values of~$n$, we can use known small sorting networks
to obtain somewhat reasonable constructions.

\section{Spectating Is Hard}

Imagine we turn the tables and, instead of considering the magician's challenge
in hanging a picture on $n$ nails with certain properties, we consider the
spectator's challenge of choosing which nails to remove.  A natural objective,
if the spectator is shy and wants to get off stage as quickly as possible,
is to remove as few nails as possible in order to make the picture fall.
Unfortunately for the spectator, for a given picture hanging,
this problem is NP-complete and hard to approximate:

\begin{theorem} \label{hardness theorem}
  For a given picture hanging on $n$ nails, it is NP-complete to decide
  whether there are $k$ nails whose removal fells the picture,
  and it is hard to approximate the minimum number of nails within an
  $\epsilon \log n$ factor for some $\epsilon > 0$.
\end{theorem}

The decision problem is in NP: given the free-group representation of a
picture hanging as input, and which $k$ nails to remove as certificate,
we can verify that the word cancels after removing the necessary nails.

For hardness, we reduce from the Set Cover problem:
given a universe $U=\{u_1,u_2,\ldots,u_m\}$ of $m$ elements,
and a collection of $n$ subsets, ${\cal S}=\{S_1,S_2,\ldots,S_n\}$,
where each $S_i \subseteq U$, find a smallest subcollection
${\cal S}' \subseteq {\cal S}$ whose union $\bigcup_{S\in{\cal S}'} S = U$.
This problem is NP-hard to approximate within an $\epsilon' \log n$ factor.

Now, given an instance of Set Cover, we create a picture hanging on $n$ nails
as follows.  Each $x_i$ corresponds to a set~$S_i$.
Let $E_j$ denote the efficient $1$-out-of-$n$ construction
from Section~\ref{Polynomial construction} applied to the $x_i$'s
corresponding to $S_j$'s that contain~$u_j$.
We construct $E_j$ for each $j$ between $1$ and $m$,
and then combine them with a balanced tree of $\AND$'s
according to Equation~(\ref{and eq}).
This picture hanging falls precisely when every element has at least 
one containing set chosen for removal,
which means that the sets were all covered.
The objective is the same for the two problems, and the problem size
increased by only a polynomial factor,
by Theorems~\ref{basic} and~\ref{arbitrary theorem}.

This argument completes the proof of Theorem~\ref{hardness theorem}.

\medskip

We can similarly argue that it is NP-hard for the attention-hoarding
spectator who aims to \emph{maximize} the number of nails to remove before
felling the picture hanging.  By the same reduction, this problem becomes
finding a set of elements that hit every set in the collection $\cal S$,
which is the Hitting Set problem.  Reversing the roles of elements and
sets, we have the identical Set Cover problem.  Inapproximability no longer
follows because the objectives are reversed.

\section{Open Problems}
\label{open}

Several interesting open questions remain
about the optimality of our constructions:

\begin{enumerate}

\item
Does the $1$-out-of-$n$ picture hanging puzzle require a solution of
length $\Omega(n^2)$?

\item
What is the complexity of finding the shortest picture hanging
for a given monotone Boolean function?

\item
In the reverse direction, does a short solution to a picture-hanging puzzle
imply a low-depth monotone circuit for the monotone Boolean formula of the
puzzle?


\item
For the spectator, is there an $O(\log n)$-approximation algorithm
for removing the fewest nails to fell the picture hanging?

\item
How does the complexity of picture hanging change when allowing the rope
to twist around itself, not just the nails?  Michael Paterson suggested
this problem, and points out that that the $1$-out-of-$n$ puzzle admits
a linear-complexity solution in this model, as shown in
Figure~\ref{knotted rope}.

\begin{figure}
  \centering
  \includegraphics[width=\linewidth]{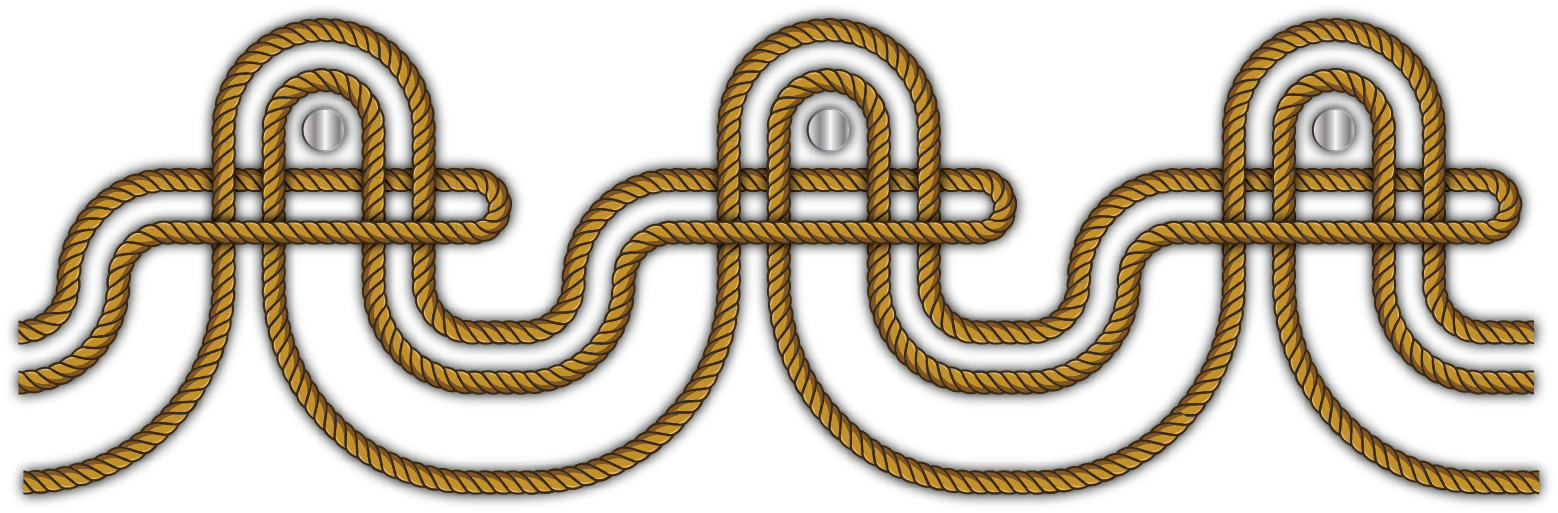}
  \caption{Knotting the rope with itself, and not just the nails,
           enables a $1$-out-of-$n$ construction using $O(n)$ crossings.
           The left and right edges of the diagram wrap around.
           (Construction by Michael Paterson, June 2012.)}
  \label{knotted rope}
\end{figure}

\end{enumerate}

\iffull
\section*{Acknowledgments}
\else
\textbf{Acknowledgments.}
\fi
We thank Jason Cantarella for helpful early discussions,
Kim Whittlesey for pointing out reference~\cite{Stanford},
and the anonymous referees for helpful comments.

\let\realbibitem=\bibitem
\def\bibitem{\par \vspace{-0.9ex}\realbibitem}

\bibliography{picturehanging}
\bibliographystyle{alpha}

\appendix
\section{Puzzle Solutions}
\label{solutions}

This appendix gives solutions to the puzzles from Section \ref{puzzles}
in the free-group notation defined in Section~\ref{free group}.
These solutions are based on the basic theory described in
Section~\ref{theory}, though they do not all fall under the specific
statement of Theorem~\ref{disjoint theorem}.
The solutions are not unique; shorter solutions may well exist.

\solution{\ref{1outta3}}
Figure \ref{1outta3 fig} shows
$S_3 =
\x_1 \x_2 \x_3 \NOT{\x_2} \NOT{\x_3} \NOT{\x_1} \x_3 \x_2 \NOT{\x_3} \NOT{\x_2}$.

\solution{\ref{2outta3}}
$\x_1 \x_2 \x_3 \NOT{\x_1} \NOT{\x_2} \NOT{\x_3}$.
\xxx{draw this}

\solution{\ref{1or2outta3}}
$[x_1, x_2 x_3] = \x_1 \x_2 \x_3 \NOT{\x_1} \NOT{\x_3} \NOT{\x_2}$,
where the nails are ordered from left to right.

\solution{\ref{1outta4}}
$E(1:4) =
\x_1 \x_2 \NOT{\x_1} \NOT{\x_2} \x_3 \x_4 \NOT{\x_3} \NOT{\x_4} \x_2 \x_1 \NOT{\x_2} \NOT{\x_1} \x_4 \x_3 \NOT{\x_4} \NOT{\x_3}$.

\solution{\ref{2outta4}}
$[[x_1 x_2, [x_1 x_3, x_1 x_4]], [x_2 x_3, [x_2 x_4, x_3 x_4]]] =
\x_1 \x_2 \x_1 \x_3 \x_1 \x_4 \NOT{\x_3} \NOT{\x_1} \NOT{\x_4} \NOT{\x_1} \NOT{\x_2} \NOT{\x_1} \x_1 \x_4 \x_1 \x_3 \NOT{\x_4} \NOT{\x_1} \NOT{\x_3} \NOT{\x_1} \x_2 \x_3 \x_2 \x_4 \x_3 \x_4 \NOT{\x_4} \NOT{\x_2} \NOT{\x_4} \NOT{\x_3} \NOT{\x_3} \NOT{\x_2} \x_3 \x_4 \x_2 \x_4 \NOT{\x_4} \NOT{\x_3} \NOT{\x_4} \NOT{\x_2} \x_1 \x_3 \x_1 \x_4 \NOT{\x_3} \NOT{\x_1} \NOT{\x_4} \NOT{\x_1} \x_1 \x_2 \x_1 \x_4 \x_1 \x_3 \NOT{\x_4} \NOT{\x_1} \NOT{\x_3} \NOT{\x_1} \NOT{\x_2} \NOT{\x_1} \x_2 \x_4 \x_3 \x_4 \NOT{\x_4} \NOT{\x_2} \NOT{\x_4} \NOT{\x_3} \x_2 \x_3 \x_3 \x_4 \x_2 \x_4 \NOT{\x_4} \NOT{\x_3} \NOT{\x_4} \NOT{\x_2} \NOT{\x_3} \NOT{\x_2}$.

\solution{\ref{3outta4}}
$\x_1 \x_2 \x_3 \x_4 \NOT{\x_1} \NOT{\x_2} \NOT{\x_3} \NOT{\x_4}$.

\solution{\ref{2or2outta4}}
$[\x_1 \x_2, \x_3 \x_4] =
\x_1 \x_2 \x_3 \x_4 \NOT{\x_2} \NOT{\x_1} \NOT{\x_4} \NOT{\x_3}$,
where $x_1$ and $x_2$ are red and $x_3$ and $x_4$ are blue.
\xxx{draw this}

\solution{\ref{1or2outta4}}
$[S_2, \x_3 \x_4] = \x_1 \x_2 \NOT{\x_1} \NOT{\x_2} \x_3 \x_4
                    \x_2 \x_1 \NOT{\x_2} \NOT{\x_1} \NOT{\x_4} \NOT{\x_3}$,
where $x_1$ and $x_2$ are red and $x_3$ and $x_4$ are blue.

\solution{\ref{1or3outta6}}
$[S_3, x_4 x_5 x_6] =
\x_1 \x_2 \x_3 \NOT{\x_2} \NOT{\x_3} \NOT{\x_1} \x_3 \x_2 \NOT{\x_3} \NOT{\x_2} \x_4 \x_5 \x_6 \x_2 \x_3 \NOT{\x_2} \NOT{\x_3} \x_1 \x_3 \x_2 \NOT{\x_3} \NOT{\x_2} \NOT{\x_1} \NOT{\x_6} \NOT{\x_5} \NOT{\x_4}$,
where $x_1$, $x_2$, and $x_3$ are red and $x_4$, $x_5$, and $x_6$ are blue.

\solution{\ref{1or2outta6}}
$[S_3, x_4 x_5 x_6 \NOT{x_4} \NOT{x_5} \NOT{x_6}] =
\x_1 \x_2 \x_3 \NOT{\x_2} \NOT{\x_3} \NOT{\x_1} \x_3 \x_2 \NOT{\x_3} \NOT{\x_2} \x_4 \x_5 \x_6 \NOT{\x_4} \NOT{\x_5} \NOT{\x_6} \x_2 \x_3 \NOT{\x_2} \NOT{\x_3} \x_1 \x_3 \x_2 \NOT{\x_3} \NOT{\x_2} \NOT{\x_1} \x_6 \x_5 \x_4 \NOT{\x_6} \NOT{\x_5} \NOT{\x_4}$,
where $x_1$, $x_2$, and $x_3$ are red and $x_4$, $x_5$, and $x_6$ are blue.

\solution{\ref{two1soutta6}}
$[[[\x_1\x_3, [\x_2\x_4, \x_1\x_5]], [\x_3\x_6, [\x_1\x_4, \x_2\x_3]]], 
  [[\x_1\x_6, [\x_2\x_5, \x_4\x_6]], [\x_3\x_5, [\x_2\x_6, \x_4\x_5]]]] =
\x_1 \x_3 \x_2 \x_4 \x_1 \x_5 \NOT{\x_4} \NOT{\x_2} \NOT{\x_5} \NOT{\x_1} \NOT{\x_3} \NOT{\x_1} \x_1 \x_5 \x_2 \x_4 \NOT{\x_5} \NOT{\x_1} \NOT{\x_4} \NOT{\x_2} \x_3 \x_6 \x_1 \x_4 \x_2 \x_3 \NOT{\x_4} \NOT{\x_1} \NOT{\x_3} \NOT{\x_2} \NOT{\x_6} \NOT{\x_3} \x_2 \x_3 \x_1 \x_4 \NOT{\x_3} \NOT{\x_2} \NOT{\x_4} \NOT{\x_1} \x_2 \x_4 \x_1 \x_5 \NOT{\x_4} \NOT{\x_2} \NOT{\x_5} \NOT{\x_1} \x_1 \x_3 \x_1 \x_5 \x_2 \x_4 \NOT{\x_5} \NOT{\x_1} \NOT{\x_4} \NOT{\x_2} \NOT{\x_3} \NOT{\x_1} \x_1 \x_4 \x_2 \x_3 \NOT{\x_4} \NOT{\x_1} \NOT{\x_3} \NOT{\x_2} \x_3 \x_6 \x_2 \x_3 \x_1 \x_4 \NOT{\x_3} \NOT{\x_2} \NOT{\x_4} \NOT{\x_1} \NOT{\x_6} \NOT{\x_3} \x_1 \x_6 \x_2 \x_5 \x_4 \x_6 \NOT{\x_5} \NOT{\x_2} \NOT{\x_6} \NOT{\x_4} \NOT{\x_6} \NOT{\x_1} \x_4 \x_6 \x_2 \x_5 \NOT{\x_6} \NOT{\x_4} \NOT{\x_5} \NOT{\x_2} \x_3 \x_5 \x_2 \x_6 \x_4 \x_5 \NOT{\x_6} \NOT{\x_2} \NOT{\x_5} \NOT{\x_4} \NOT{\x_5} \NOT{\x_3} \x_4 \x_5 \x_2 \x_6 \NOT{\x_5} \NOT{\x_4} \NOT{\x_6} \NOT{\x_2} \x_2 \x_5 \x_4 \x_6 \NOT{\x_5} \NOT{\x_2} \NOT{\x_6} \NOT{\x_4} \x_1 \x_6 \x_4 \x_6 \x_2 \x_5 \NOT{\x_6} \NOT{\x_4} \NOT{\x_5} \NOT{\x_2} \NOT{\x_6} \NOT{\x_1} \x_2 \x_6 \x_4 \x_5 \NOT{\x_6} \NOT{\x_2} \NOT{\x_5} \NOT{\x_4} \x_3 \x_5 \x_4 \x_5 \x_2 \x_6 \NOT{\x_5} \NOT{\x_4} \NOT{\x_6} \NOT{\x_2} \NOT{\x_5} \NOT{\x_3} \x_3 \x_6 \x_1 \x_4 \x_2 \x_3 \NOT{\x_4} \NOT{\x_1} \NOT{\x_3} \NOT{\x_2} \NOT{\x_6} \NOT{\x_3} \x_2 \x_3 \x_1 \x_4 \NOT{\x_3} \NOT{\x_2} \NOT{\x_4} \NOT{\x_1} \x_1 \x_3 \x_2 \x_4 \x_1 \x_5 \NOT{\x_4} \NOT{\x_2} \NOT{\x_5} \NOT{\x_1} \NOT{\x_3} \NOT{\x_1} \x_1 \x_5 \x_2 \x_4 \NOT{\x_5} \NOT{\x_1} \NOT{\x_4} \NOT{\x_2} \x_1 \x_4 \x_2 \x_3 \NOT{\x_4} \NOT{\x_1} \NOT{\x_3} \NOT{\x_2} \x_3 \x_6 \x_2 \x_3 \x_1 \x_4 \NOT{\x_3} \NOT{\x_2} \NOT{\x_4} \NOT{\x_1} \NOT{\x_6} \NOT{\x_3} \x_2 \x_4 \x_1 \x_5 \NOT{\x_4} \NOT{\x_2} \NOT{\x_5} \NOT{\x_1} \x_1 \x_3 \x_1 \x_5 \x_2 \x_4 \NOT{\x_5} \NOT{\x_1} \NOT{\x_4} \NOT{\x_2} \NOT{\x_3} \NOT{\x_1} \x_3 \x_5 \x_2 \x_6 \x_4 \x_5 \NOT{\x_6} \NOT{\x_2} \NOT{\x_5} \NOT{\x_4} \NOT{\x_5} \NOT{\x_3} \x_4 \x_5 \x_2 \x_6 \NOT{\x_5} \NOT{\x_4} \NOT{\x_6} \NOT{\x_2} \x_1 \x_6 \x_2 \x_5 \x_4 \x_6 \NOT{\x_5} \NOT{\x_2} \NOT{\x_6} \NOT{\x_4} \NOT{\x_6} \NOT{\x_1} \x_4 \x_6 \x_2 \x_5 \NOT{\x_6} \NOT{\x_4} \NOT{\x_5} \NOT{\x_2} \x_2 \x_6 \x_4 \x_5 \NOT{\x_6} \NOT{\x_2} \NOT{\x_5} \NOT{\x_4} \x_3 \x_5 \x_4 \x_5 \x_2 \x_6 \NOT{\x_5} \NOT{\x_4} \NOT{\x_6} \NOT{\x_2} \NOT{\x_5} \NOT{\x_3} \x_2 \x_5 \x_4 \x_6 \NOT{\x_5} \NOT{\x_2} \NOT{\x_6} \NOT{\x_4} \x_1 \x_6 \x_4 \x_6 \x_2 \x_5 \NOT{\x_6} \NOT{\x_4} \NOT{\x_5} \NOT{\x_2} \NOT{\x_6} \NOT{\x_1}$,
where $x_1$ and $x_2$ are red, $x_3$ and $x_4$ are green, and
$x_5$ and $x_6$ are blue.

\end{document}